\documentclass[a4paper]{VisionStyle}
\usepackage{epsfig}

\begin{document}
\title{A New Look at the X-ray Properties of Our Galaxy}

\author{R.S.\,Warwick\inst{}} 

\institute{
  Department of Physics and Astronomy, University of Leicester, 
  Leicester LE1 7RH, UK }

\maketitle 

\begin{abstract}

Observational programmes currently underway with both {\em Chandra} and 
{\em XMM-Newton} are set to revolutionize our view of our own Galaxy. This
is fortunate since the X-ray band can provide crucial diagnostics in the 
quest to understand the composition and structure of the Galaxy and the 
high energy processes which influence its evolution. Here we review 
several areas in which rapid progress is being made. These include
investigations into the nature of the Galactic X-ray source population at 
faint fluxes, the distribution and origin of the X-ray emission 
emanating from the Galactic Centre Region and the spectral characteristics 
of the Galactic Soft X-ray Background.

\end{abstract}

\section{Introduction}
  
Viewed from outside, say from a distance of a few Mpc, our own Galaxy would 
appear as a fairly unremarkable, but multi-faceted X-ray source.  
Table~\ref{rwarwick-E1_tab:tab1} provides a very rough breakdown of the 
contribution that the various X-ray emitting objects, populations and 
diffuse components make to the total X-ray luminosity of the Milky Way. 
The brightest few dozen LMXRB/HMXRB dominate the integrated discrete 
source signal, with diffuse 
emission associated with the Galactic Disk, Bulge and Halo providing a 
roughly matching contribution. At the present, the active galactic nucleus 
 is not a particularly significant source of X-ray emission
(\cite{rwarwick-E1:Bag01}), although there are some indications of
episodes of much more luminous activity in the recent past 
(\cite{rwarwick-E1:Koy96}; \cite{rwarwick-E1:Mur01}; 
\cite{rwarwick-E1:Mae02}).

\begin{table}[bht]
  \caption{X-ray emitting populations in the Milky Way. }
  \label{rwarwick-E1_tab:tab1}
  \begin{center}
    \leavevmode
    \footnotesize
    \begin{tabular}[h]{lcr}
      \hline \\[-5pt]
      Type of Emitter & Number in   &  Summed $\rm L_X$ \\
                      & Galaxy & $10^{38} \rm~erg~s^{-1}$ \\[+5pt]
      \hline \\[-5pt]
      HMXRB  & $\sim 30$  &   $\sim 3$ \\
      LMXRB  & $\sim 100$ &   $\sim 30$ \\
      SNR    & $\sim 500$ &   $ <1$ \\
      CVs  & $\sim 10^{4}$ &  $ <1$ \\
      RSCVn& $\sim 10^{6}$ &  $ <1$ \\
      Stellar Coronae & $\sim 10^{10}$ & $ <1$ \\
	&	&	\\
      Active Nucleus  & 1 & $ <0.001 $ \\
	&	&	\\
      Diffuse Disk & 1  & $ \sim 3$ \\
      Diffuse Bulge & 1 & $ \sim 20$ \\
      Diffuse Halo  & 1 & $ \sim 10$ \\
	&	&	\\
      Total & & $\sim 60$ \\
      \hline \\
      \end{tabular}
  \end{center}
\end{table}

The spatial distribution and luminosity function of the brightest discrete
sources have recently been investigated by \cite*{rwarwick-E1:Gri01}. These 
authors also
demonstrate that the Galaxy's X-ray binary population is considerably more 
luminous than that in M31, but unexceptional against the yardstick of more 
active 
spirals and giant elliptical galaxies. Unfortunately the spatial extent and 
luminosity of the diffuse emission, particularly that associated with the 
Galactic halo, is much less well defined and we have to look for Milky Way 
analogues amongst nearby galaxies to gain a more complete perspective
(e.g. \cite{rwarwick-E1:Wan02a}).

In this paper, we will not be looking at the most X-ray luminous 
components of our Galaxy but, instead, focusing on three areas
of Galactic X-ray Astronomy in which major advances can be expected
as a result of on-going {\em XMM-Newton} and {\em Chandra}  programmes.
These two missions provide, essentially for the first time,
instruments capable of carrying out high sensitivity, coherent surveys of 
selected 
regions of our Galaxy. Although, of course, {\em ROSAT} has conducted a 
definitive soft X-ray all-sky survey, for Galactic studies access to
the hard 2--10 keV band is paramount in order to overcome the
obscuration of the cool interstellar medium. For example, towards
the Galactic Centre photoelectric absorption in a hydrogen column density 
of $\sim 6 \times 10^{22} \rm~cm^{-2}$  (\cite{rwarwick-E1:Bag01}) gives
rise to a sharp cut-off in X-ray spectra below $\sim 2.5$ keV.

\begin{figure*}[htb]
\begin{center}
\epsfig{file=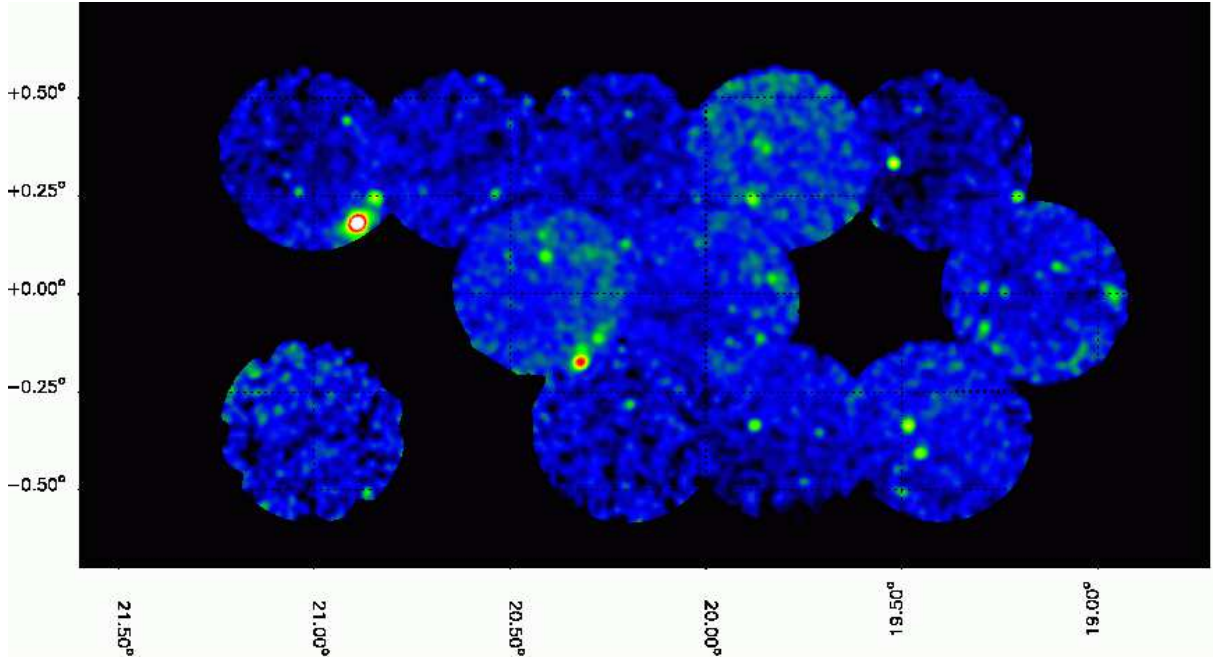,angle=270,width=16cm}
\end{center}
\caption{A mosaic of the MOS1+2 images from 12 XGPS fields. The energy
range is 2--6 keV. Only a bright subset of sources are visible in this
representation. To date the survey covers a narrow (1 \degr wide) strip in the
Galactic longitude range $19\degr - 21.5 \degr$.}  
\label{rwarwick-E1_fig:fig1}
\end{figure*}

\begin{figure*}[htb]
\begin{center}
\epsfig{file=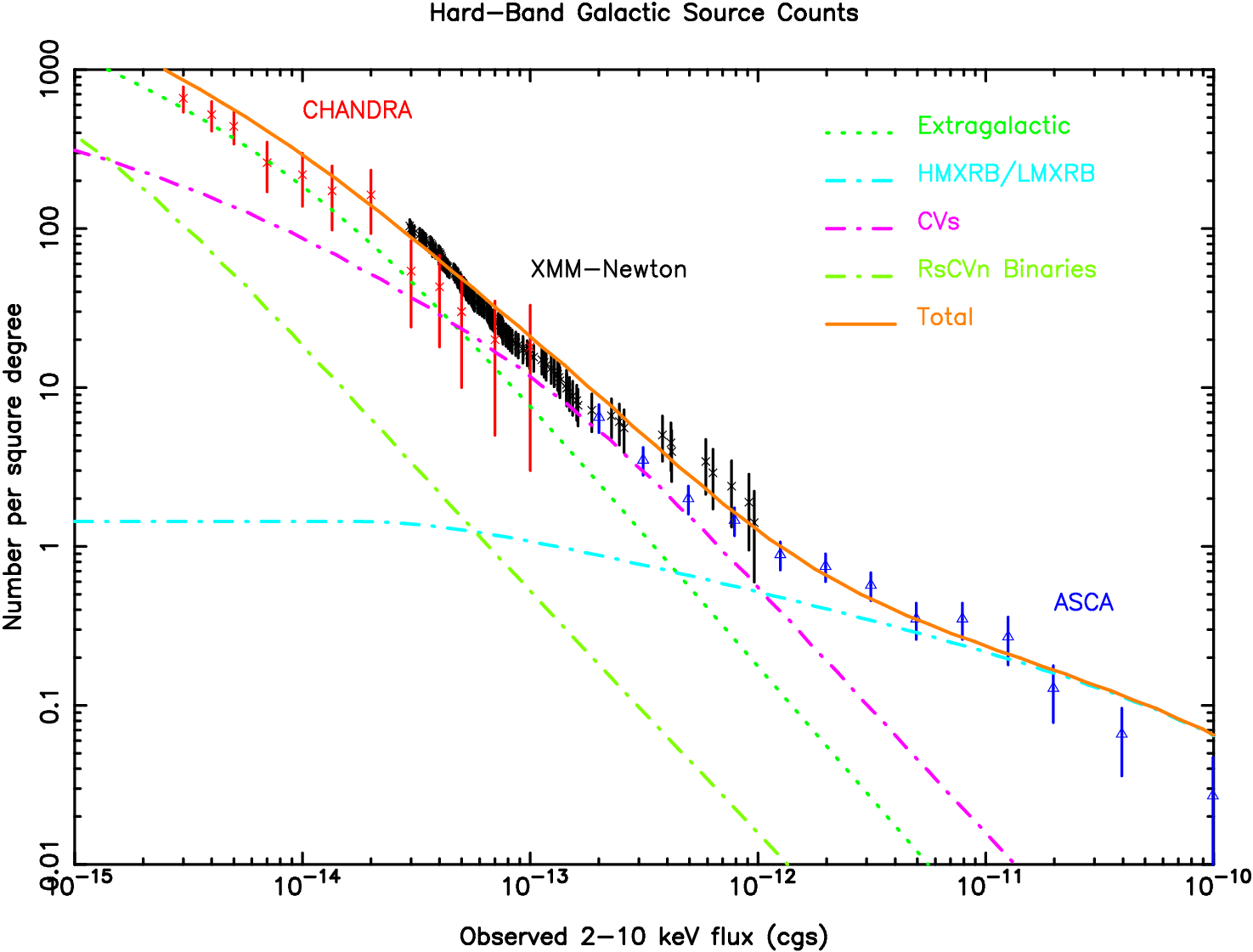,angle=270,width=16cm}
\end{center}
\caption{The 2--10 keV log N - log S relation measured in the Galactic 
Plane based on 
{\em XMM-Newton}, {\em Chandra} and {\em ASCA} observations.  The curves 
show the predicted source counts for various Galactic source
populations including intermediate luminosity X-ray binaries, cataclysmic
variables and RS CVn binaries. The predicted contribution of extragalactic 
sources is also shown.
}  
\label{rwarwick-E1_fig:fig2}
\end{figure*}

\section{Faint Galactic X-ray Source Populations}

Both {\em XMM-Newton} and  {\em Chandra} are carrying out surveys in 
the Galactic Plane. One such programme is the {\em XMM-Newton} Galactic  Plane 
Survey (hereafter the XGPS survey) which aims to map a narrow 5\degr~strip 
of the plane near longitude 22\degr. A total of 40 short observations 
amounting to 200 ks exposure time were awarded for this purpose in AO-1.  
To date about 40\% of the XGPS observations have been carried out;
Fig. \ref{rwarwick-E1_fig:fig1} shows a mosaic of 12 of the XGPS fields.

So far a total of 223  discrete sources have been detected in the XGPS survey
at a significance  greater than 5$\sigma$ in either the EPIC-pn and/or 
EPIC-MOS cameras (for more details, see \cite{rwarwick-E1:Han02}).
We have used the XGPS source catalogue to construct a log N - log S 
curve for the low Galactic latitude sky. The normalisation and slope of this 
relation can, in principle, provide important information  on the  spatial 
distribution and luminosity function of the various Galactic source 
populations, albeit bound-up with line-of-sight absorption effects.  
Fig. \ref{rwarwick-E1_fig:fig2} shows the derived 2--10 keV log N - log S 
distribution for the XGPS sources after correcting for sky coverage effects. 
For comparison the source counts derived from the extensive  
survey of the Galactic plane carried by {\em ASCA} 
(\cite{rwarwick-E1:Sug01}) and from recent deep {\em Chandra}  
(\cite{rwarwick-E1:Ebi01}) observations are also shown. 
As can be seen from the figure, the flux range probed
by the XGPS  survey is intermediate between that sampled by the
{\em ASCA} and {\em Chandra}  programmes. The slope of  the 
log N - log S curve is, as expected, flatter than that measured for the 
extragalactic sky.

\begin{figure}[htb]
\begin{center}
\epsfig{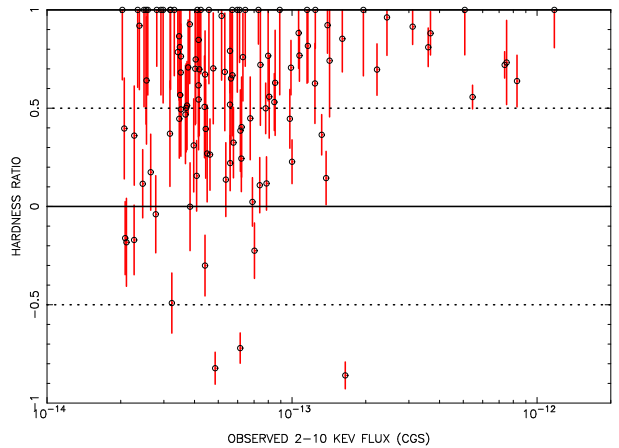}
\end{center}
\caption{Hardness ratio versus 2-10 keV flux for the XGPS sources.
The hardness ratio is defined as $(M-S)/M+S)$ where S is the measured 
0.5--2 keV count rate and M the corresponding 2--6 keV rate.}  
\label{rwarwick-E1_fig:fig3}
\end{figure}

The measured Galactic source counts have been compared with predictions 
for various source populations based on fairly simple source distribution 
and luminosity models (Fig. \ref{rwarwick-E1_fig:fig2}). This preliminary
analysis suggests that the Galactic population dominates down to a flux of 
$5 \times 10^{-14}$ $\rm~erg~s^{-1}~cm^{-2}$ (2--10 keV) below which the 
contribution of extragalactic sources (seen through the appreciable
Galactic column density in this direction) grows rapidly. The hardness ratio 
distribution of the sources (Fig. \ref{rwarwick-E1_fig:fig3}) exhibits 
significant scatter but is consistent with the view that
a highly absorbed extragalactic population emerges largely at the 
lower end of flux range sampled in the XGPS survey. This analysis 
confirms that the strategy of the XGPS survey, namely
relatively short (5--10 ks) observations but fairly wide-angle coverage,
is well tuned to the objective of studying Galactic X-ray source 
populations at faint fluxes. 

The XGPS survey has  so far  yielded in excess of 200 point
source detections and  a realistic target for the full AO1 programme is  
an X-ray source catalogue with between 500  and 1000 entries. This will 
provide a valuable resource for studying the Galactic
X-ray source population at faint fluxes ({\it i.e.} down to
$F_X \sim 2 \times 10^{-14} \rm~erg~s^{-1}~cm^{-2}$). The challenge is 
then to conduct an intensive programme of optical identifications
so as to confirm the nature of the underlying populations.
Work in this area is currently underway (e.g. \cite{rwarwick-E1:Mot02}).

\section{X-rays from the Galactic Centre Region}
\label{rwarwick-E1_sec:gc}

\begin{figure*}[htb]
\begin{center}
\hbox{
\epsfig{file=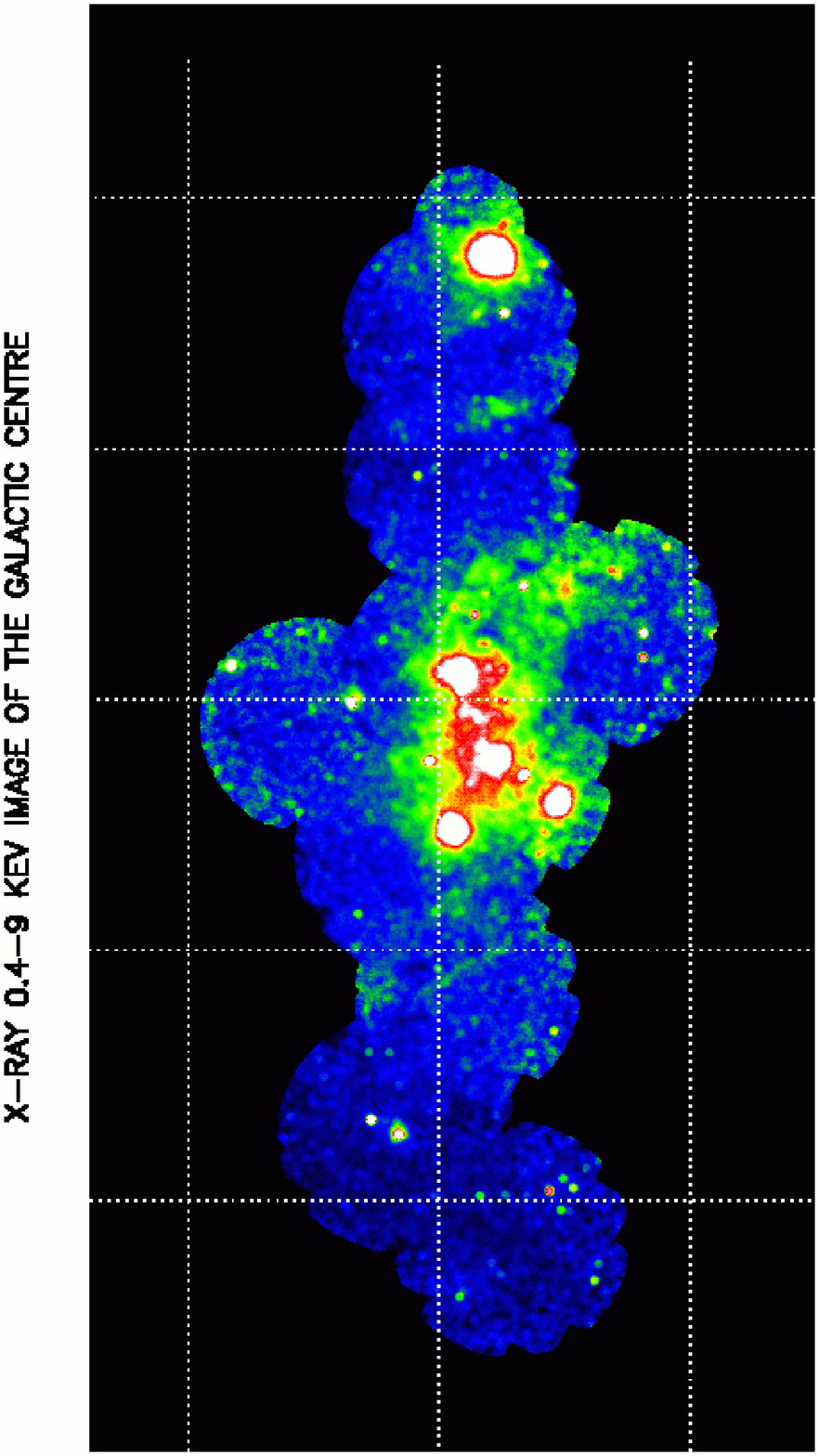,angle=270,width=15.8cm}
}
\vspace{0.5cm}
\hbox{
\epsfig{file=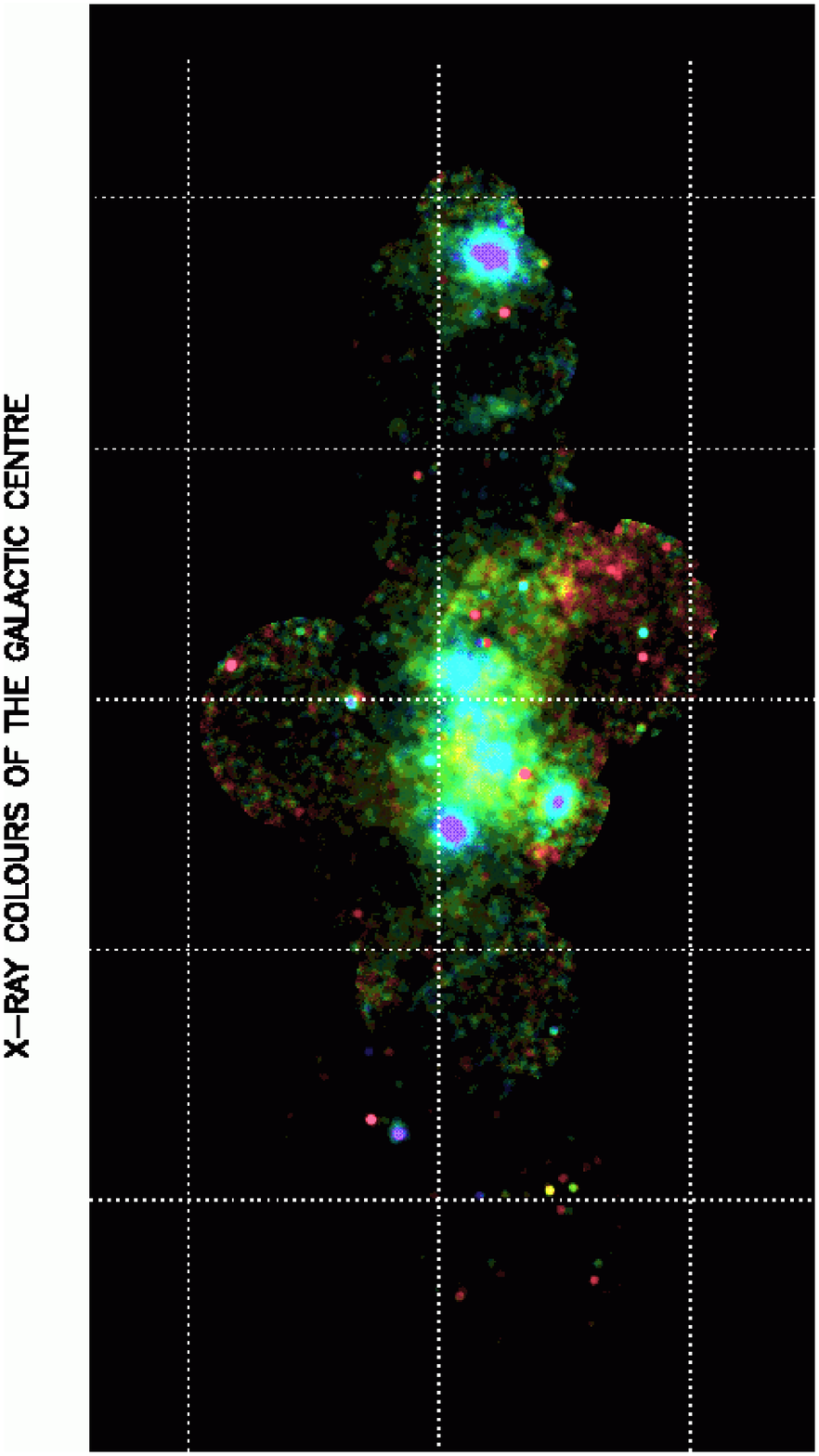,angle=270,width=15.8cm}
}
\end{center}
\caption{Top panel: A mosaic of the MOS1+2 images covering the region
around the Galactic Centre. The energy range is 0.4--9 keV. 
Bottom panel: A corresponding false-colour image based on three energy 
bands 0.5--2 keV (red), 2--5 keV (green), 5--9 keV (blue).
In both images the grid is drawn in Galactic Coordinates centred on 
(l,b) = (0\degr,0\degr) with a grid spacing of 0.5\degr. 
}  
\label{rwarwick-E1_fig:fig4}
\end{figure*}

\begin{figure*}[ht]
\begin{center}
\epsfig{file=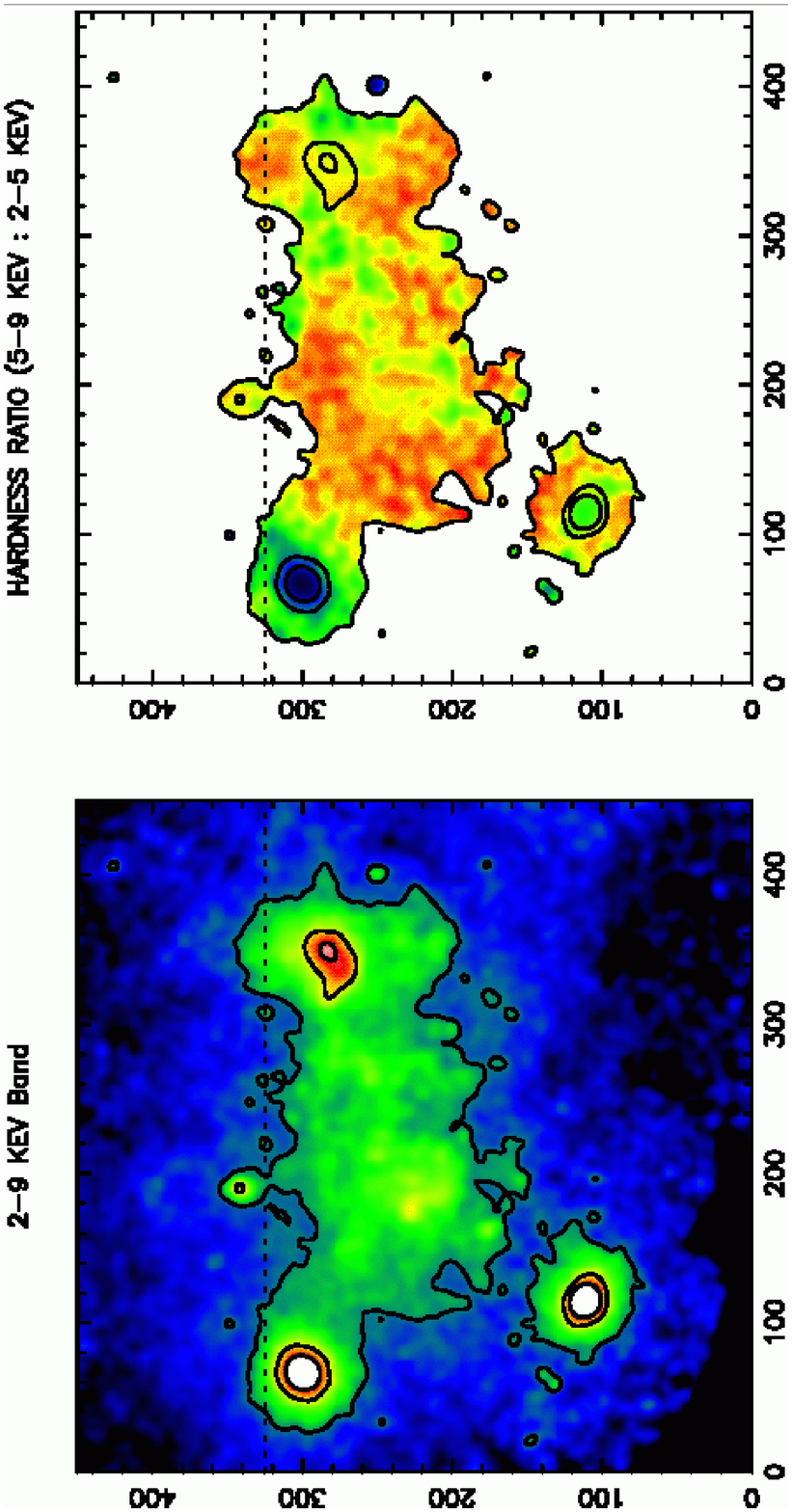,angle=270,width=15.8cm}
\end{center}
\caption{Left Panel: The distribution of 2--9 keV X-ray emission in
the Radio Arc Region. The image covers a region $0.5\degr \times
0.5 \degr$. The two bright sources at the left-hand edge of the
image are 1E1743-2843 and SAX J1740.0-2853, whereas the bright extended
source associated with Sgr A East is visible on the right-hand side of 
the image. Right Panel: The corresponding hardness (H-M/H+M) ratio 
distribution based on the hard (5--9 keV) and medium (2--5 keV) band 
images.}  
\label{rwarwick-E1_fig:fig5}
\end{figure*}

The central $\sim 300$ parsecs of our Galaxy is a unique region 
with a very complex structure which, due to the high obscuration
in the optical band, has been investigated mainly through radio, 
infrared and X-ray observations. The central
$\sim 2.6 \times 10^{6}$ solar mass, black-hole, which is spatially
coincident with the compact radio source Sgr$\rm A^{*}$
at the dynamical centre of the Galaxy, is currently in a relatively
quiescent state (\cite{rwarwick-E1:Bag01}). Nevertheless the Galactic 
Centre Region (hereafter GCR) as a whole is X-ray bright due to the presence 
of large-scale diffuse emission, several very luminous ($\rm  L_X \geq 10^{38} 
~erg~s^{-1}$) X-ray binaries and an underlying population of 
intermediate luminosity discrete sources.

A major survey of the GCR in the hard
(2--10 keV) X-ray band has previously been carried out by {\em ASCA} 
(\cite{rwarwick-E1:Koy96}; \cite{rwarwick-E1:Sak02}) and
recently the {\em Chandra} X-ray Observatory has also mapped
the region on arcsec scales (\cite{rwarwick-E1:Wan02a}; 
\cite{rwarwick-E1:Wan02b}). Here we provide a preliminary
report of a complementary {\em XMM-Newton} programme  (PI:
Anne Decourchelle) to survey
the Galactic Plane within $\pm 1\degr$ of the Galactic Centre.

The {\em XMM-Newton} programme to map the GCR
consists of 10 overlapping pointings (plus one or two additional
observations targeted at specific Galactic Centre sources). 
Fig. \ref{rwarwick-E1_fig:fig4} shows a preliminary mosaiced image
based on the MOS 1+2 datasets covering a broad 0.4--9 keV bandpass. This
image has been background-subtracted and corrected
for exposure time variations due to the telescope vignetting and 
the field overlaps. Fig. \ref{rwarwick-E1_fig:fig4} also shows a false-colour 
image based on the subdivision of the data into soft (0.4--2 keV),
medium (2--5 keV) and hard (5--9 keV) spectral bandpasses. The effects
of X-ray absorption vary across the field but will be most acute for the
soft X-ray band.

These {\em XMM-Newton} images reveal the presence of three very bright
discrete sources, namely 1E1743-2843 (at l,b = 0.3,0.0),  1E 1740.7-2942 
(at 359.1,-0.1) and SAX J1740.0-2853 (at 0.2,-0.2), together with the bright
extended X-ray emission associated with the the non-thermal radio
source Sgr A East (\cite{rwarwick-E1:Mae02}). The large-scale 
diffuse X-ray emission in the GCR has an asymmetric 
distribution, with the Radio Arc Region, 10\arcmin~to the southeast of
the Galactic Centre, exhibiting a particularly high X-ray surface brightness. 
On the other side of the Galactic Centre the diffuse X-ray emission
appears to sweep southward away from the Galactic Plane; the spectral 
softening of this component towards the edge of the field of view in Fig. 
\ref{rwarwick-E1_fig:fig4} may well be due to the reduced obscuration
off the Galactic Plane. Closer inspection of these images also reveals 
X-ray emission associated with the Sgr B2 and Sgr C complexes and several
known SNR including G0.9+0.1. An underlying  population of faint point 
sources is also apparent throughout the region.

Fig. \ref{rwarwick-E1_fig:fig5} shows in more detail  the distribution 
of diffuse emission to the east of the Galactic Centre and also the 
variation in the 5--9 keV : 2--5 keV hardness ratio across the same region. 
The source 1E1743-2843, and a second source to the west of 
Sgr $\rm A^{*}$, have particularly hard spectra. The diffuse X-ray 
component exhibits considerable spectral variations on scales
down to 1\arcmin~consistent with the complex mix of thermal and 
non-thermal components present in the Radio Arc Region 
(\cite{rwarwick-E1:LaR00}; \cite{rwarwick-E1:Bam02}). The linear 
structure which crosses Sgr A East in the hardness ratio map maybe 
indicative of a soft bipolar outflow from this region in a direction
roughly perpendicular to the Galactic Plane (see also 
\cite{rwarwick-E1:Mae02}). An alternative possibility is that there
is excess absorption in a highly flattened distribution lying roughly
parallel to the Galactic Plane such as to intercept our line
of sight to Sgr A East. This issue which will be clarified by detailed 
analysis of the available EPIC pn and MOS spectra.

\begin{figure*}[htb]
\begin{center}
\epsfig{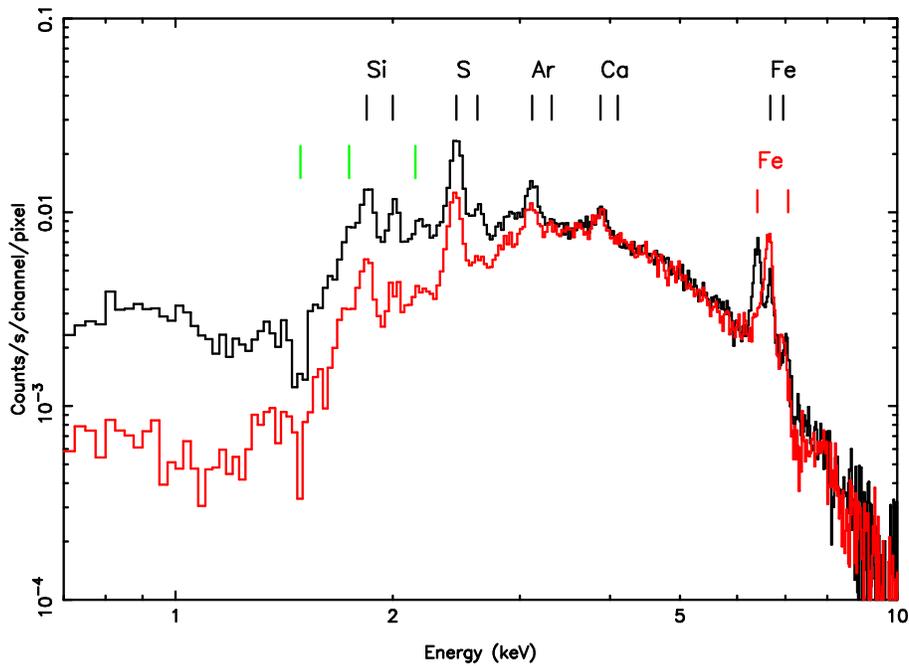}
\end{center}
\caption{A comparison of the  EPIC MOS 1+2 spectra measured for regions 
centred on Sgr A East (red) and on the brightest X-ray knots in the 
Radio Arc region.}  
\label{rwarwick-E1_fig:fig6}
\end{figure*}

To illustrate the quality of the available EPIC spectral data, we have
extracted MOS 1+2 spectra from two circular regions of radius $3'$ and
$5'$ radius, centred respectively on Sgr A East and the brightest 
knots of the Radio Arc region. Fig. \ref {rwarwick-E1_fig:fig6}
compares the resulting spectra. For these spectra the instrumental 
background has been subtracted (see next section) but no correction
has been made for energy dependence of the telescope vignetting function.
The normalisations have been scaled to align the continua in the 
4--6 keV band. 

These spectra illustrate many of the well-established properties
of the diffuse X-ray emitting components in the GCR, namely bright
K$\alpha$ lines of the helium-like and hydrogen-like ions of silicon, sulphur,
argon, calcium and iron plus a hint of some helium-like K$\beta$ lines.
The Radio Arc Region is also characterised by a bright fluorescent 
K$\alpha$ line of neutral iron, a feature which is absent from
the Sgr A East spectrum (\cite{rwarwick-E1:Mae02}). Apart from
the neutral iron line and the extra absorption which is evident
in the spectrum of Sgr A East the extracted spectra show remarkable
similarities particularly in the form of the continuum above $\sim 4$ 
keV. Why there should be such a similarity is a puzzle given
the identification of Sgr A East with a supernova remnant
(SNR 000.0+00.0),  whereas the Radio Arc region contains a complex 
blend of thermal emission structures, non-thermal filaments, SNRs,
dense clouds and star clusters.  The origin and nature of the diffuse X-ray
emission seen in the GCR and in Galactic Plane in the form of the
Galactic X-ray Ridge remains a topic of active debate, with many
of the observed properties, including the constancy of the spectrum,
casting doubt on a purely thermal origin for the emission
(\cite{rwarwick-E1:Val00}; \cite{rwarwick-E1:Tan00}). In this respect,
future detailed characterisation and analysis of the EPIC spectra from the 
{\em XMM-Newton} Galactic Centre programme will be of considerable
interest.
 
\section{The Spectrum of the Soft X-ray Background}
\label{rwarwick-E1_sec:spect}

\begin{figure*}[htb]
\begin{center}
\epsfig{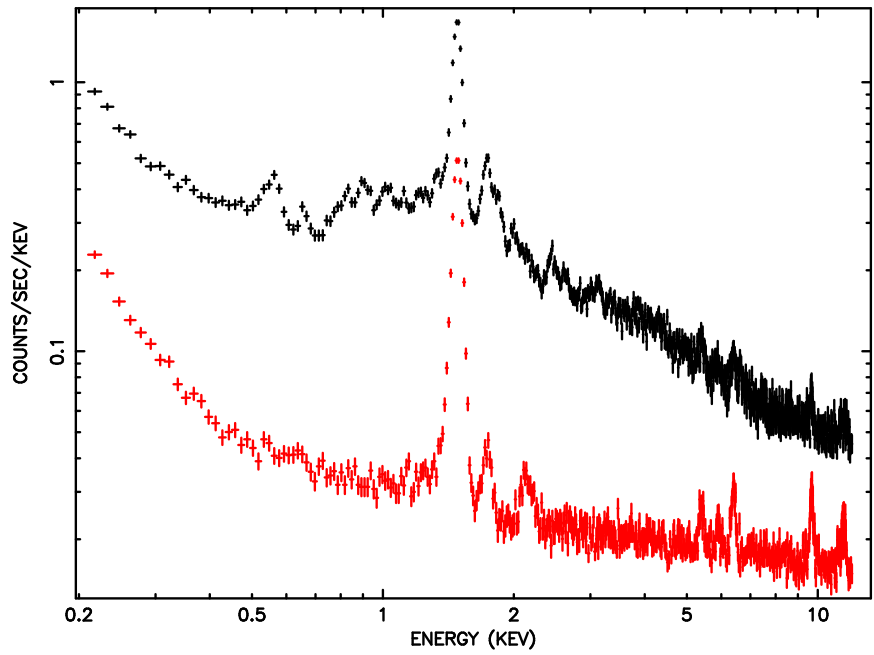}
\end{center}
\caption{The X-ray spectrum measured in the MOS cameras within the central
field of view (black) compared to that recorded in the ``unexposed'' 
edge regions of the CCDs (red).}  
\label{rwarwick-E1_fig:fig7}
\end{figure*}

\begin{figure*}[htb]
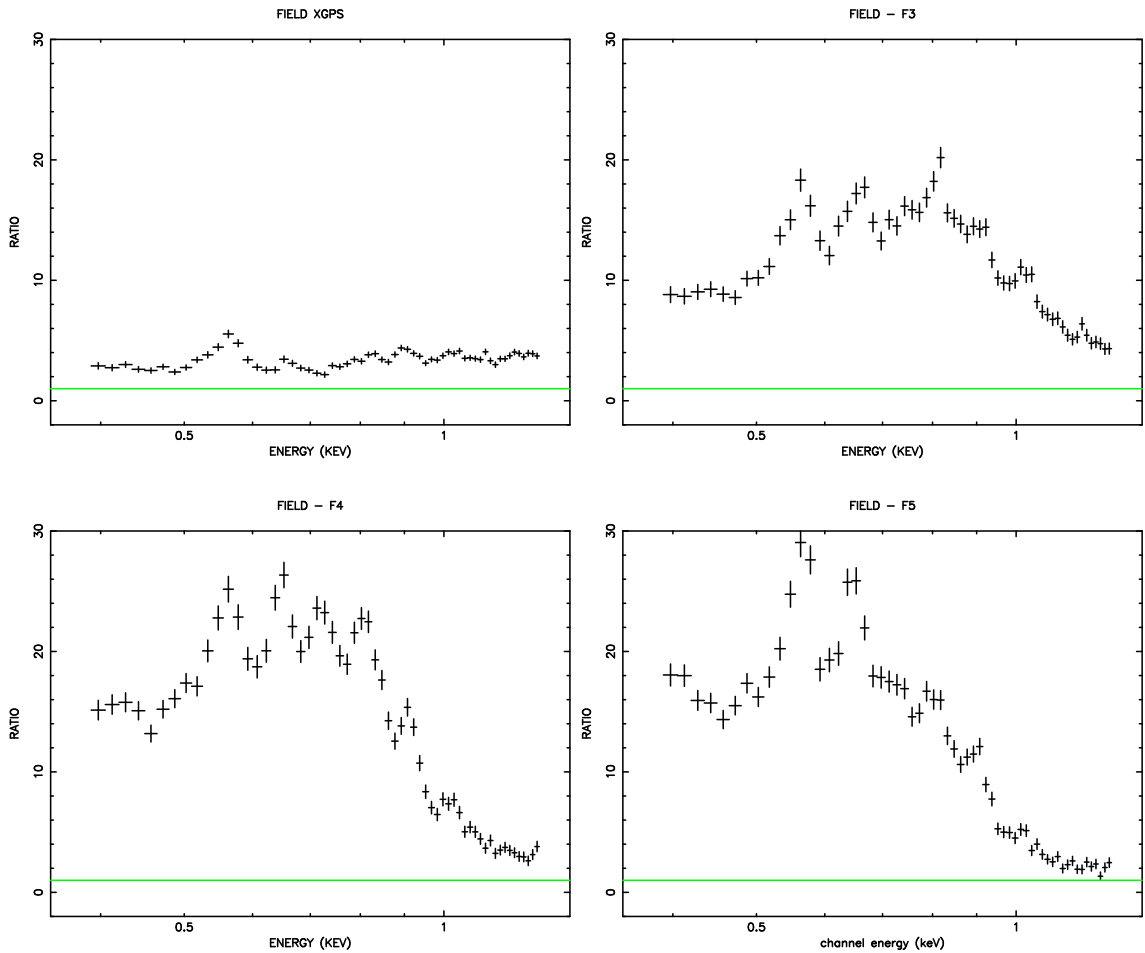

\begin{center}
\hbox{
\epsfig{file=rwarwick-E1_fig8a.ps,angle=270,width=7.5cm}
\epsfig{file=rwarwick-E1_fig8b.ps,angle=270,width=7.5cm}
}
\vspace{0.5cm}
\hbox{
\epsfig{file=rwarwick-E1_fig8c.ps,angle=270,width=7.5cm}
\epsfig{file=rwarwick-E1_fig8d.ps,angle=270,width=7.5cm}
}
\end{center}
\caption{MOS 1+2 full-field spectra derived from 4 {\sl XMM-Newton} 
observations. The four fields are: the Galactic Plane at $l \approx 
20^{\degr}$ (top-left); the southern Galactic Bulge near
$l=0.4^{\degr}$, $b=-5.5^{\degr}$ (top-right); two regions in the 
North Polar Spur (bottom panels). The data are shown ratioed to a 
power-law component with spectral index $\Gamma=1.4$.}  
\label{rwarwick-E1_fig:fig8}
\end{figure*}

The diffuse soft X-ray background (SXRB) originates predominately in the
thermal emission of  hot plasma located in the Galactic Halo, in the Galactic 
Bulge, in the Galactic Plane (forming the so-called Galactic Ridge) and in 
large loop and ring structures, such as radio Loop 1, associated with nearby 
SNRs. The ROSAT all-sky survey has provided excellent maps 
(\cite{rwarwick-E1:Sno97}) of the SXRB which delineate the various 
emission features and also reveal marked spectral variations
across the sky. The impact in certain regions, most notably the Galactic 
Plane, of photoelectric absorption in interstellar clouds is also
evident. The complex spatial distribution of the Galactic SXRB suggests 
immediately 
that it is truly diffuse emission. This is in accord with the fact that
there is no known population of discrete source which could account for 
the observed X-rays. The emission undoubtedly comes from volumes of hot 
interstellar gas but the detailed 3-d morphology, the origin of the gas, 
the heating mechanism and the physical state and composition of the material 
is very poorly understood. 

In order to study the true distribution and state of the hot gas that 
comprises the SXRB, mapping
on an angular scale of $\sim 1\degr$~ in individual spectral lines or blends 
of lines is required. A limited number of moderate 
energy resolution measurements indicate a wide range of spectral signatures 
(e.g. \cite{rwarwick-E1:Roc84}; \cite{rwarwick-E1:Gen95}; 
\cite{rwarwick-E1:Men01}). For example,
there is clear evidence that the spectrum differs strongly between the
North Polar Spur (part of Loop I) and the Galactic Bulge emission.
However, many of the available measurements are from sounding rocket 
observations using large fields of view (100's of square degrees), which
average over very significant spatial structure in order to gather
sufficient counts for spectral measurements. In contrast {\em ASCA} 
observations have a small field of view, but in practice must also be 
averaged over different directions to achieve adequate 
statistics (as well as having limited usefulness below $\sim 0.7$ keV).  
In summary,  our present understanding of the spectral composition
of the SXRB is very incomplete.

The EPIC instrument on {\em XMM-Newton} is well suited to the detection of 
extended sources with relatively low X-ray surface brightness. Following 
recent progress in characterising the instrumental background, it appears
that the spectrum of diffuse components, filling the full (30\arcmin) field of 
view of the EPIC cameras, can also be measured with good precision
(see \cite{rwarwick-E1:Lum02}). Here we report recent progress in
measuring the spectrum of the SXRB with {\em XMM-Newton}.

For this study we use the EPIC MOS cameras which have a substantial active 
CCD area outside  the nominal field of view defined  by the optical 
blocking filter.  Starting from the calibrated event lists generated by the
Science Analysis System (SAS), the full-field MOS light curves
are screened  for ``soft proton flares'', namely a highly variable
flux of low  energy  particles trapped within the magnetosphere which  
are focused  by  the mirror  systems  onto  the CCD detectors.  
In the selected ``quiescent background'' intervals, the charged particle 
induced events which are not rejected by the on-board or ground  processing 
give rise to a residual background component that is relatively  
constant in  spectrum and which shows little variation across the detector.  
The covered edge regions of the MOS CCDs provide a direct 
measure of this internal instrumental background, which can be used to
background-subtract the central field signal (Fig. 
\ref{rwarwick-E1_fig:fig7}). An extra complication is that the
the passage of charged particles through the cameras induces fluorescent 
X-ray emission in the form of emission lines at energies characteristic 
of the materials used in the camera construction. Thus, for example, 
strong K$\alpha$ lines at 1.48 keV and 1.74 keV due, respectively, to 
aluminium  in the camera body and silicon in the CCDs are present in the 
MOS spectra (Fig. \ref{rwarwick-E1_fig:fig7}). Unfortunately, these 
fluorescent lines are not uniformly distributed across the MOS CCDs 
(\cite{rwarwick-E1:Lum02}) and  it is necessary to exclude regions of the 
spectrum which, even after background-subtraction, remain
contaminated by such features.

Our preliminary measurements of the spectrum of the SXRB are shown in Fig. 
\ref{rwarwick-E1_fig:fig8}. The four fields sample the Galactic Plane,
the inner Galactic Bulge and two locations in the North Polar Spur.
Emission lines due to OVII, OVIII, FeXVII, NeIX, NeX and Mg XI are detected 
in three out of the four fields with only OVII 
apparent, presumably as a foreground component, in the Galactic Plane.
The spectra vary considerable from field to field
reflecting changes both in the temperature, composition and possibly
the degree of ionization non-equilibrium of the plasma. 
Also, the Galactic Plane and Bulge regions are characterized by 
hard components that are significantly more intense than in either of 
the Loop 1 fields.  As a check we have compared  the SXRB surface brightness
derived from MOS spectra in the 0.5--2.0 keV band with 
that measured in the diffuse SXRB maps from the RASS.
The correlation is very good, verifying that the spectra in
Fig. \ref{rwarwick-E1_fig:fig8} provide good estimates of the true 
SXRB signal. 

To date we have carried out only limited quantitative 
analysis of these spectra. The evidence for two
temperature components is quite strong in all fields (typically with
$kT \sim0.2$ and $\sim0.5$ keV). There is also evidence
that the abundances are significantly lower than solar, although this
will depend on the temperature structure. It may be that the higher Z material
is found preferentially  in the cold  gaseous phase of the ISM or is locked 
up in interstellar
grains. In the future we will undertake 
a much more detailed analysis, for example the ratio of the OVIII/OVII 
measures the ionization conditions and, for isochoric cooling with 
$T>10^{6.25}$ K, provides a good indicator of the temperature. The Fe line 
emission at around 0.7 keV is also sensitive to non-equilibrium ionization 
effects.

\section{Conclusions}
\label{rwarwick-E1_sec:conc}

The {\em XMM-Newton} and {\em Chandra} missions are proving to be
the drivers of a revolution in our knowledge of the high energy properties 
of our Galaxy.  This paper has focussed on the spectral imaging capabilities 
of {\em XMM-Newton} which are very well suited to the study of 
faint Galactic source  populations, active regions of the Galaxy such as the 
GCR and the low-surface brightness features of the diffuse SXRB.

\begin{acknowledgements}

This paper refers to a number of preliminary results from on-going 
{\em XMM-Newton}  programmes. The author would like to thank many colleagues
and, in particular, A. Decourchelle, A. Hands, D. Helfand, D. Lumb, M. Sakano, 
M. Watson and R. Willingale, for allowing unpublished results
from collaborative work to be included in this paper.

\end{acknowledgements}


\begin{thebibliography}{}

\bibitem[\protect\astroncite{Baganoff et~al.}{2001}]{rwarwick-E1:Bag01}
Baganoff F. et~al., 2001, Nature, 413, 45 

\bibitem[\protect\astroncite{Bamba et~al.}{2002}]{rwarwick-E1:Bam02}
Bamba A. et~al., 2002, these proceedings 

\bibitem[\protect\astroncite{Ebisawa et~al.}{2001}]{rwarwick-E1:Ebi01}
Ebisawa K., Maeda Y., Kaneda H., Yamauchi S., 2001, Science 293, 1633

\bibitem[\protect\astroncite{Gendreau et~al.}{1995}]{rwarwick-E1:Gen95}
Gendreau K.C., et~al., 1995, PASJ, 47, L5

\bibitem[\protect\astroncite{Grimm et al.}{2001}]{rwarwick-E1:Gri01}
Grimm L-H., Gilfanov M., Sunyaev R.,  2001, astro.ph/0109239

\bibitem[\protect\astroncite{Hands et~al.}{2002}]{rwarwick-E1:Han02}
Hands A, Warwick R., Watson M., Helfand D., 2002, these proceedings

\bibitem[\protect\astroncite{Koyama et~al.}{1996}]{rwarwick-E1:Koy96}
Koyama K., Maeda Y., Sonobe T., Takeshima T., Tanaka Y., Yamauchi S.,
1996, PASJ, 48, 249

\bibitem[\protect\astroncite{LaRosa et~al.}{2000}]{rwarwick-E1:LaR00}
LaRosa T.N., Kassim N.E., Lazio T.J.W., Hyman S.D., 2000, ApJ, 119, 207
 
\bibitem[\protect\astroncite{Lumb et~al.}{2002}]{rwarwick-E1:Lum02}
Lumb D.H., Warwick R.S., Page M., de Luca, A., 2002, A\&A, submitted

\bibitem[\protect\astroncite{Maeda et~al.}{2002}]{rwarwick-E1:Mae02}
Maeda Y., et~al., 2002, ApJ, in press

\bibitem[\protect\astroncite{Mendenhall \& Burrows}{2001}]{rwarwick-E1:Men01}
Mendenhall J.A., Burrows D.N. 2001, ApJ, 563, 716

\bibitem[\protect\astroncite{Motch et~al.}{2002}]{rwarwick-E1:Mot02}
Motch C., et~al., 2002, these proceedings

\bibitem[\protect\astroncite{Murakami et~al.}{2001}]{rwarwick-E1:Mur01}
Murakami H., Koyama K., Maeda Y., 2001, ApJ, 558, 687

\bibitem[\protect\astroncite{Rocchia et~al.}{1984}]{rwarwick-E1:Roc84}
Rocchia R., et~al., 1984, A\&A, 130, 53

\bibitem[\protect\astroncite{Sakano et~al.}{2002}]{rwarwick-E1:Sak02}
Sakano M., et~al., 2002, ApJS, 138, 19

\bibitem[\protect\astroncite{Snowden et~al.}{1997}]{rwarwick-E1:Sno97}
Snowden S.L., et~al. 1997, ApJ, 485, 125

\bibitem[\protect\astroncite{Sugizaki et~al.}{2001}]{rwarwick-E1:Sug01}
Sugizaki M., et al., 2001, ApJS, 134, 77

\bibitem[\protect\astroncite{Tanaka et~al.}{2000}]{rwarwick-E1:Tan00}
Tanaka Y., et al., 2000, PASJ, 52, L25

\bibitem[\protect\astroncite{Valinia et~al.}{2000}]{rwarwick-E1:Val00}
Valinia A., et al., 2000, ApJ, 543, 733

\bibitem[\protect\astroncite{Wang}{2002a}]{rwarwick-E1:Wan02a}
Wang Q.D., 2002, these proceedings

\bibitem[\protect\astroncite{Wang et~al.}{2002b}]{rwarwick-E1:Wan02b}
Wang Q.D., Gotthelf, E.V., Lang C.C., 2002, Nature, 415, 148

\end{thebibliography}
\end{document}